\documentclass[twocolumn,superscriptaddress]{revtex4-1}
\usepackage{bm}
\usepackage{epsfig}
\usepackage{amsmath}
\usepackage{color}
\newcommand{\nix}[1]{}
\usepackage[unicode=true,colorlinks=true,urlcolor=blue,citecolor=blue]{hyperref}
\usepackage{graphicx}

\usepackage{eurosym}
\usepackage{xcolor}
\usepackage{ulem}
\usepackage{soul}

\begin{document}
	
\title{Elastic electron scattering and localization in a chain with isotopic disorder}

\author{K. S. Denisov}
\affiliation{Ioffe Institute, 194021 St. Petersburg, Russia}
\affiliation{Department of Physics, University at Buffalo, State University of New York, Buffalo, NY 14260, USA}

\author{E. Ya. Sherman}
\affiliation{Department of Physical Chemistry, University of the Basque Country, 48940,
Leioa, Spain}
\affiliation{Ikerbasque, Basque Foundation for Science, 48009, Bilbao, Spain}
\affiliation{EHU Quantum Center, University of the Basque Country UPV/EHU, 48940 Leioa, Bizkaia, Spain}

\begin{abstract}
We study elastic electron scattering and localization by ubiquitous isotopic disorder in one-dimensional systems
appearing due to interaction with phonon modes localized at isotope impurities. By using a tight-binding model with
intersite hopping matrix element dependent on the interatomic distance, 
we find mass-dependent backscattering  probability by single and pairs of isotopic 
impurities. For the pairs, in addition to the mass, the distance between the isotopes plays the 
critical role. Single impurities effectively attract electrons and can produce localized weakly bound electron states.
In the presence of disorder, the electron free path at positive energies becomes finite and the corresponding 
Anderson localization at the spatial scale greatly exceeding the distance between 
the impurities becomes possible. 
\end{abstract}

\date{\today}
\maketitle

\section{Introduction}

Ubiquitous isotopic disorder leads to a finite zero-temperature resistivity 
even in metals where all other kinds of disorder do not exist. 
Since isotopic substitution does not produce an explicitly position-dependent random
potential, the understanding and analysis of this residual resistivity is a highly nontrivial problem. 
This purely quantum effect based on the zero point atomic motion 
was understood in Refs. \cite{Pomeranchuk,Kagan}. {The approaches of Ref. \cite{Pomeranchuk}
and Ref. \cite{Kagan} are different. Reference \cite{Pomeranchuk} considered random kinetic energy of the lattice 
vibrations as the source of the electron scattering in terms of higher-order, beyond 
the first Born approximation, scattering theory. Later, it was shown in Ref. \cite{Kagan} that 
the Born approximation can be applied taking into account randomization of the crystal lattice Debye-Waller factor 
by isotopic disorder.} Both approaches result in a very small nonzero zero-temperature resistivity. 
Being the origin of unusual elastic and inelastic electron scattering \cite{Vandecasteele}, 
the isotopic disorder calls for studies 
of electron localization, usually considered as a result of a static randomly position-dependent potential,
qualitatively different from the isotopic disorder {producing a random field of dynamical finite frequency phonon modes.} 
Here we study these effects in a one-dimensional system \cite{Berezinskii}
with isotopic disorder, where all the scattering processes can be presented in a clear explicit form. 

Various aspects of the role of lattice vibrations for quantum transport have been studied for a long time 
(see, e.g., Refs. \cite{Kumar,Rammer,Bergmann,Glazov,Gogolin}). These approaches have been concentrated
either on the zero-point motion of impurities \cite{Kumar,Rammer,Bergmann} or on the 
effects of phonons \cite{Glazov,Gogolin} without taking into account the isotopic disorder.
Here we consider the aspect of this problem related to isotopic disorder specific 
for zero point quantum motion in localized vibrational modes.

The problem of electron interaction with isotopic disorder is directly related to 
the quantum impurity physics, where the electron interaction with internal quantum structure of the impurity-related states 
strongly influences the electron kinetics. This interaction leads to inelastic scattering  
processes, as studied, e.g., in Ref. \cite{Borda}. For isotopic defects
the internal spectrum of the impurity is the localized vibrational 
mode. As we show below, this quantum effect can lead to elastic scattering eventually resulting in electron localization.

This paper is organized as follows. In Sec. II we briefly present already known results on localized phonon modes and electron-phonon 
coupling in the form required for the analysis of electron scattering by isotopic impurities. 
In Sec. III we formulate the scattering problem, identify corresponding 
virtual phonon process, and study elastic scattering by two configurations of isotopic impurities. 
Next, we show how the localization occurs and demonstrate its characteristic features. 
Conclusions and relation to other results will be given in Sec. IV.

\section{Localized modes and electron-phonon coupling}
	
\subsection{Eigenmodes with isotopic disorder} 

Vibrational eigenmodes in crystals of various dimensionalities with isotopic disorder 
have been studied for a long time and are 
well understood by now \cite{Dyson,Lifshitz1956,Lifshitz1965,Maradudin1958,Maradudin_book,Lifshitz_book}.
It is worth noting that several advanced numerical approaches \cite{Protik,Mondal} for these modes have been 
proposed recently to supplement mainly  the analytical calculations of 
Refs. \cite{Dyson,Lifshitz1956,Lifshitz1965,Maradudin1958,Maradudin_book,Lifshitz_book}.

\begin{figure}[h]
	\includegraphics*[width=0.45\textwidth]{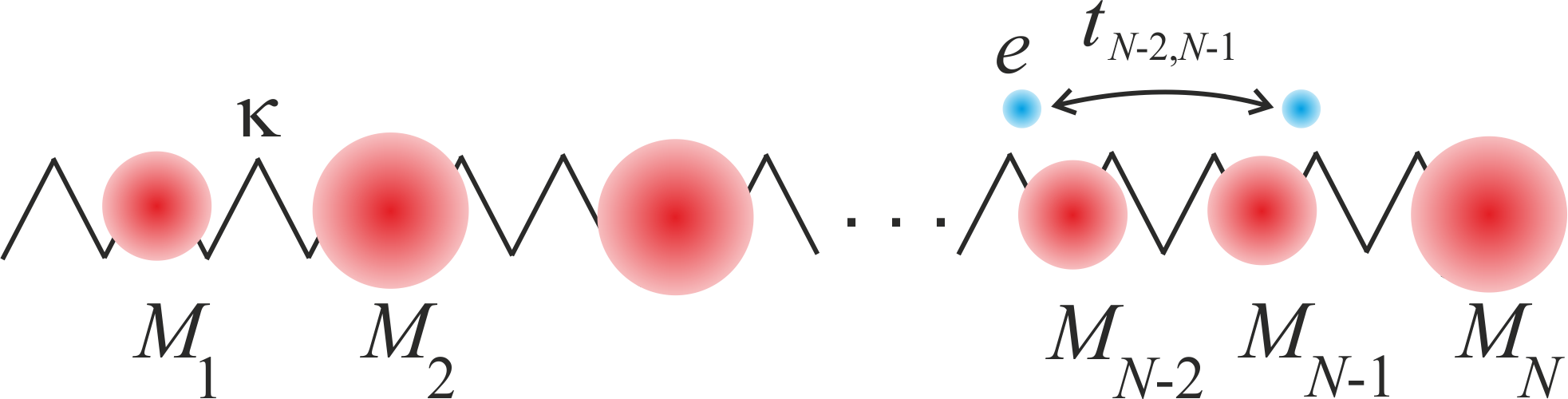}\\
	\caption{Chain with isotopic disorder (cf. Eq. (\ref{matrices})). Here $\kappa$ is the 
		elastic force constant, and $t_{n,n+1}$ is the electron bond-dependent hopping between two neighboring sites.} 
\label{chain}
\end{figure}

To provide a background for the following analysis of electron localization,
we begin by presenting and analyzing some known results for an isotopically disordered chain  
with $N\gg\,1$ atoms (see Fig. \ref{chain}) described by the Lagrangian: 
\begin{equation}
\mathcal{L} = \frac{1}{2} \left(
{\dot{\mathbf u}}^{T}\hat{M}{\dot{\mathbf u}} - \mathbf{u}^{T}\hat{K}\mathbf{u}
\right), 
\end{equation}
where ${\mathbf u}$ is $N-$component vector of atomic displacements, and matrices
\begin{equation}\label{matrices}
\hat{M} = \begin{pmatrix}
	M_{1} & 0 & 0 & \dots\\
	0 & M_{2} & 0 & \dots\\
	0 & 0 & M_{3} & \dots\\
	\dots & \dots & \dots & \dots\\
\end{pmatrix},
\quad
\hat{K} = \begin{pmatrix}
	2 \kappa & - \kappa & 0 & \dots\\
	- \kappa & 2 \kappa & - \kappa & \dots\\
	0 & - \kappa &2 \kappa & \dots\\
	\dots & \dots & \dots & \dots\\
\end{pmatrix}
\end{equation}
define the positions of isotopes with masses $M_{n}$ at the $n$th site,
and elastic force constants, respectively. For calculations
with the Born - von Karman periodic boundary conditions one has $\hat{K}_{1,N}=\hat{K}_{N,1}=-\kappa.$

The characteristic equation $ {\rm det}\left[\hat{K} - \hat{M}\omega^{2} \right]=0$  
yields the corresponding $\left\{\omega_{\alpha}^{2},\phi_{n}^{\alpha}\right\},$
where $\alpha = 1,\ldots,N$ numerates the eigenmodes with eigenvectors $\phi_{n}^{\alpha}$ 
and statistical properties of the frequencies $\omega_{\alpha}^{2}-$distribution 
obtained in the pioneering paper of Dyson \cite{Dyson}. 
For the isotopically clean system we get $\omega^{2}(k) = \omega_{\rm B}^{2}\sin^{2}{k a_0 /2}$
($\omega_{\rm B} = 2\sqrt{\kappa/M}$ is the phonon frequency at the Brillouin zone boundary,
and $a_{0}$ is the lattice constant of the chain) 
with $\phi_{k}(n) = e^{i k a_{n}}/\sqrt{N}$ and $\sum_{n}\phi_{k}^\ast(n)\phi_{k^\prime}(n) = \delta_{k,k^{\prime}}$,
here $a_{n} = n a_{0}$ is the position of the  $n$th site. 
The orthogonality and normalization condition reads as 
$\left(\mathbf{u}_{\alpha}^{T}\hat{M}\mathbf{u}_{\beta}\right)= \delta_{\alpha \beta}.$

The operator of the displacement vector in the second quantization form is given by
\begin{equation}
	\hat{{u}}_{n} = \sum_{\alpha} \sqrt{\frac{\hbar }{2M \omega_{\alpha}}}  
	\left( b_{\alpha}^{\dagger} + b_{\alpha} \right) \phi_{n}^{\alpha}, 
\label{unphin}	
\end{equation}
where $b_{\alpha}, b_{\alpha}^\dagger$ are the Bose annihilation and creation operators,
$M = \langle M_{n} \rangle$ is the average mass of atoms in a chain and normalized
$\phi_{n}^\alpha$ now includes dimensionless prefactor $\sqrt{M/M_{n}}$. 
To consider electron scattering by phonon modes one also needs the Fourier transformation 
\begin{equation}
	\label{eq:four-uq}
	\hat{{u}}_{q} = \left( \bm{\phi}_{q}^{\ast}\cdot\hat{\bm{u}}\right) = 
	\sum_{\alpha} \sqrt{\frac{\hbar }{2M\omega_{\alpha}}}\left(b_{\alpha}^{\dagger} + b_{\alpha} \right) \phi_{q}^{\alpha}, 
\end{equation}
where $ \phi_{q}^{\alpha}$ is obtained from eigenvectors as
\begin{equation}
	\label{eq:four}
	\phi_{q}^{\alpha} = \left(\bm{\phi}_{q}^{\ast} \cdot \bm{\phi}^{\alpha} \right) = \frac{1}{\sqrt{N}} \sum_{n} e^{- i q a_{n}} \phi_{n}^\alpha.
\end{equation} 

Here we consider a realization {with identical isotope impurities of masses $M_{i}$} being smaller 
than the mass $M$ of other atoms in the chain. 
For a single isolated isotope there exists 
one vibrational mode that splits off from the continuous spectrum of phonons to form a bound state 
localized nearby \cite{Kosevich}. For a small $|\Delta M|/M\ll\,1$ 
{(we restrict our consideration below to this realistic approximation)}
the frequency of this mode  $\omega_{\rm loc}=\omega_{\rm B}+\omega_{\rm B}(\Delta M/M)^{2}/2$ 
lies slightly above~$\omega_{\rm B}$. 

\begin{figure}[t]
	\centering
	\includegraphics[width=0.45\textwidth]{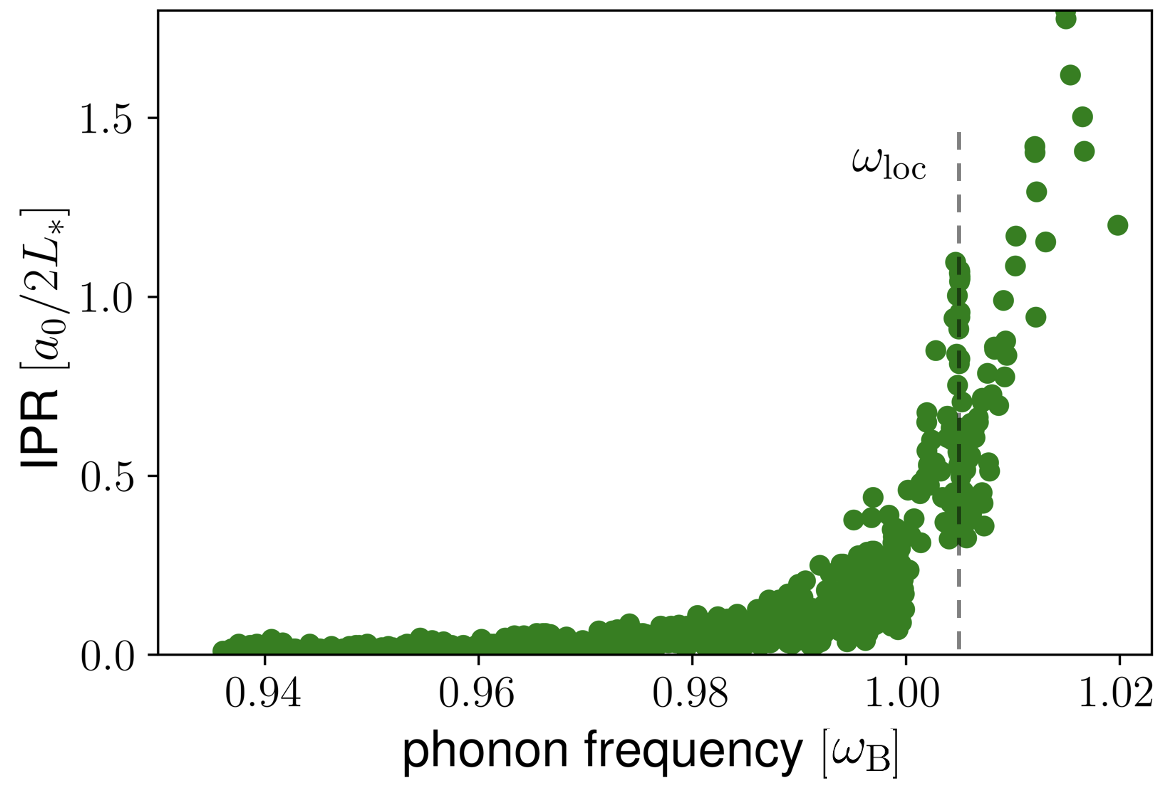}
	\caption{IPR distribution of phonon modes for $\Delta M/M = - 0.1$, $\nu_{\rm is} = 1/20a_{0}$, 
	the frequency of the localized mode $\omega_{\rm loc}/\omega_{\rm B} \approx 1.005.$} 
	\label{fig:1}
\end{figure} 

The localized mode $\phi^{\rm loc}_{n}$ centered at $n_{0}$ is described by \cite{Kosevich}
\begin{equation}
	\phi^{\rm loc}_{n} = (-1)^{n-n_{0}}\sqrt{\frac{a_{0}}{L_{\ast}}} e^{- a_{0}|n-n_{0}| / L_{\ast}}, 
	\quad
	\frac{a_{0}}{L_{\ast}} = 2 \frac{|\Delta M|}{M}.
\end{equation}
Here $L_{\ast}$ is the mode localization length. The resulting Fourier structure of the isotope-related mode being localized 
and centered at the Brillouin zone boundaries $q=\pm\,q_{\rm B},$  where $q_{\rm B} = \pi/a_{0}$, 
can be presented, 
assuming $n_{0}=0,$ in the Lorentzian form as:
\begin{eqnarray}\label{eq:eqloc}
&& \phi^{\rm loc}_{q} = \frac{2}{\sqrt{N a_0}}\frac{1}{L_{\ast}^{3/2}}\times \\
&& \left(\frac{1}{(q-q_{\rm B})^{2} + L_{\ast}^{-2}} + \frac{1}{(q+q_{\rm B})^{2} + L_{\ast}^{-2}} \right). \notag
\end{eqnarray}
Since $\phi^{\rm loc}_{n} =\phi_{-n}^{\rm loc}$ is symmetric, the Fourier component $\phi^{\rm loc}_{q}= \phi^{\rm loc}_{-q}$ is real. 
The Fourier distribution of the localized mode has the width $L_{\ast}^{-1}$.

To characterize spatial localization of the modes we use the inverse participation ratio (IPR)
\begin{equation}
\mathcal{I}^{\alpha}=\sum_{n} 
\left(\phi_{n}^\alpha\right)^{4}, 
\end{equation}
where for delocalized modes 
$\mathcal{I}^{\alpha}=O(1/N)$. 
A single isotope impurity produces a localized mode with 
$\mathcal{I}^{\rm loc}\approx a_{0}/2L_{\ast}.$  Other $N-1$ phonons are delocalized with the corresponding 
$\mathcal{I}^{\alpha}\ll\,a_{0}/L_{\ast}$ and do not have a smooth Fourier structure. 
In Fig. \ref{fig:1} we present the $\mathcal{I}^{\alpha}$
for diluted disordered system where the mean distance between the light isotopes  
$1/\nu_{\rm is}=4L_{\ast}.$ 
This Figure shows a clear distinction between localized and delocalized modes. 
It is interesting to mention that the mapping of phonon localization \cite{Dorokhov,Dorokhov2} 
on the Anderson localization \cite{Anderson} and the Green's function analysis \cite{Nieuwenhuizen} similar to \cite{Halperin}
show that at a finite concentration of impurities low-frequency phonon modes are localized, albeit
with a vanishing $\mathcal{I}^{\alpha}$ in the zero-energy limit (see Ref. \cite{Hodges} for review).

Also, Fig.~\ref{fig:1} indicates the presence of states with frequencies $\omega > \omega_{\rm loc}$ and 
larger $\mathcal{I}^\alpha$, suggesting smaller localization lengths. 
These states arise from isotopes located at the distances of the order of or less than $L_{\ast}.$ 
Here  we consider two identical isotope impurities and for a qualitative analysis 
present the displacement as $u_{n} = (-1)^{n}u(x) $
and use a continual model \cite{Kosevich} for the envelope function $u(x)$ with impurities presented as
$\delta-$potentials at distance $a:$
\begin{eqnarray}\label{eq:twoimp1}
	&& s^{2}u^{\prime\prime}(x)-a_{0}\frac{\Delta M}{M}\omega_{\rm B}^{2}\left[\delta(x-a/2)+\delta(x+a/2)\right]u(x) = \notag\\
	&&2\omega_{\rm B}(\omega-\omega_{\rm B})u(x),
\end{eqnarray}
where $s=\omega_{\rm B}a_{0}/2$ is the speed of sound, resulting in
the equation for $\omega$ 
\begin{equation}\label{eq:twoimp2}
\left(2\xi-D\right)e^{\xi a}= {\pm} D, 
\end{equation}
with $\xi\equiv \sqrt{2\omega_{\rm B}(\omega-\omega_{\rm B})}/s$ and 
$D\equiv -a_{0}\omega_{\rm B}^{2}{\Delta M}/Ms^{2}.$

\begin{figure}[t]
	\centering
	\includegraphics[width=.44\textwidth]{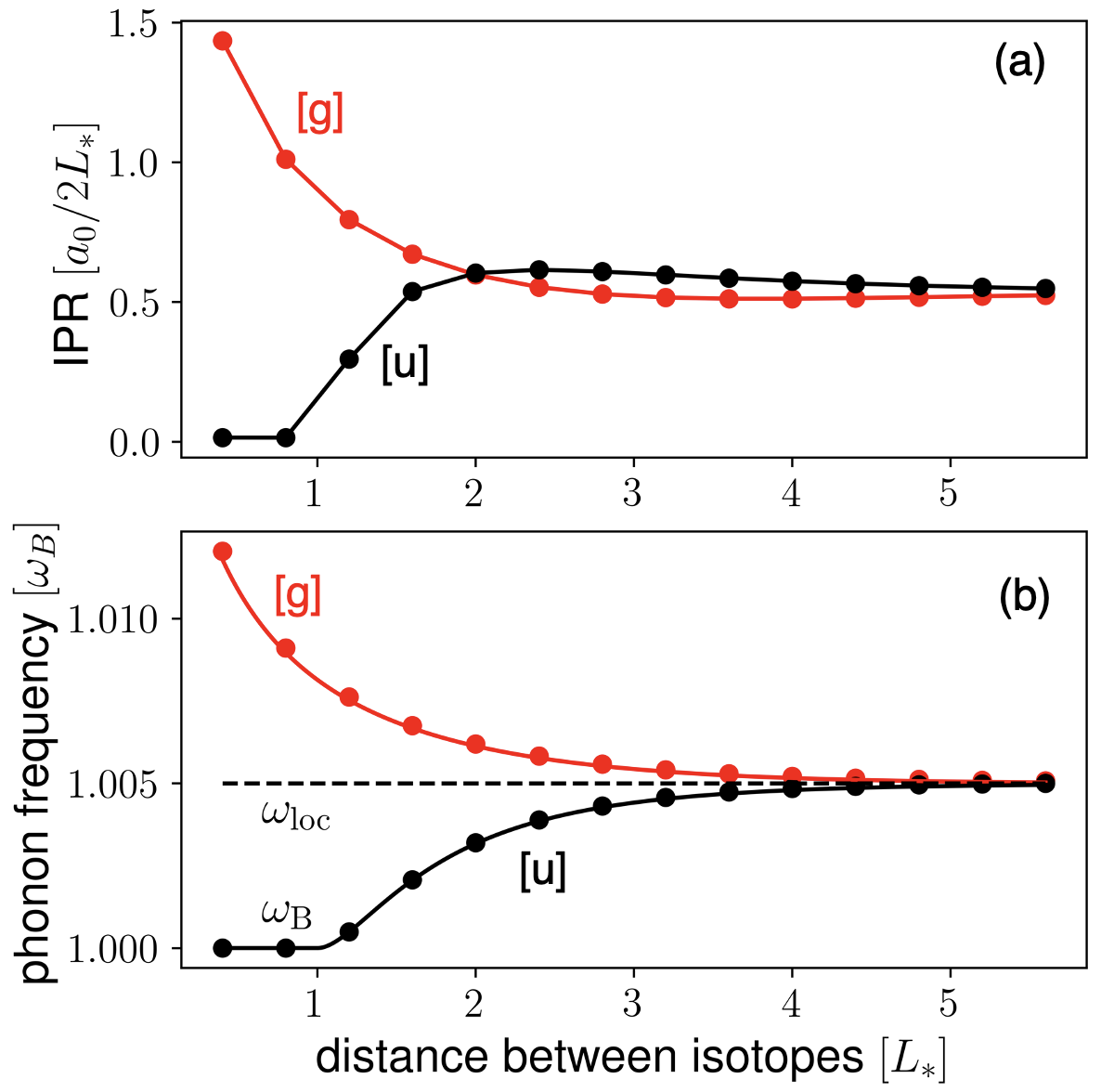}\\
	\caption{(a) The  IPR  and (b) the frequency of two highest frequency modes produced by
a pair of isotopes as a function of the distance between them,  $\Delta M/M = - 0.1$.} 
	\label{fig:IPR-2}
\end{figure}

\begin{figure}[h]
	\includegraphics[width=.5\textwidth]{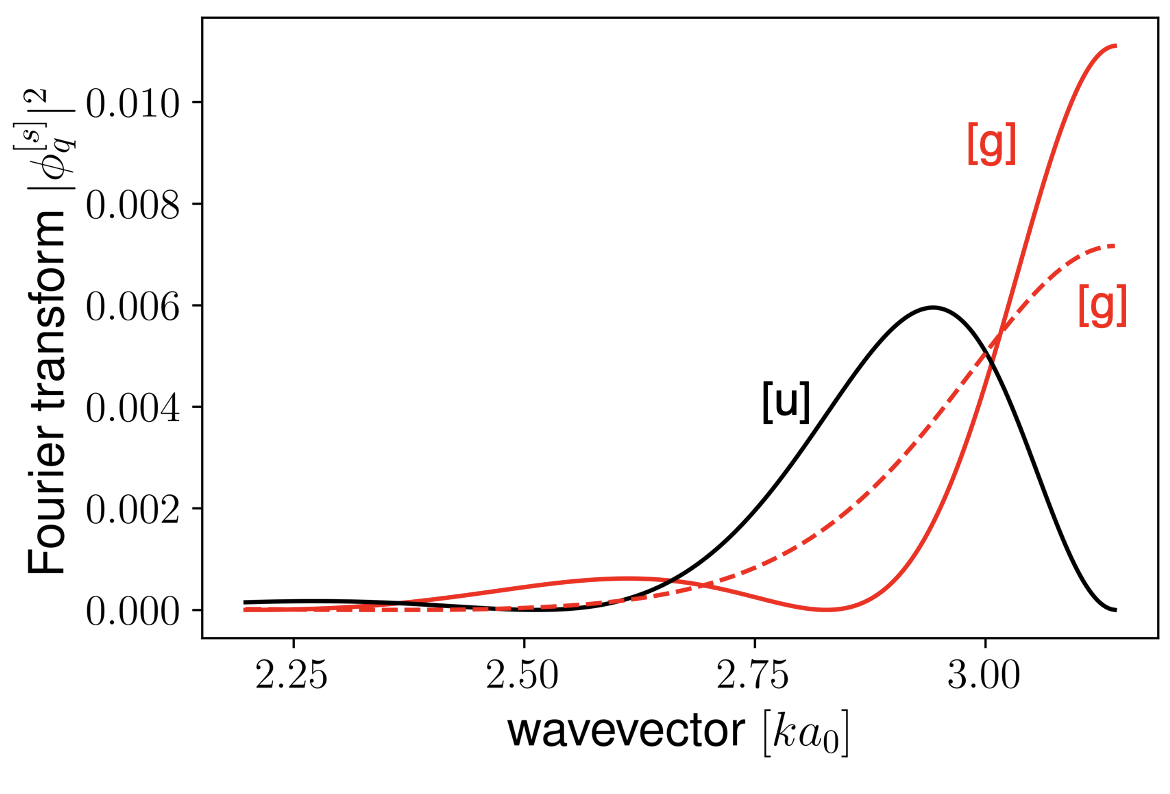}\\
	\caption{Fourier $|\phi_{q}^{[s]}|^{2}$ for odd and even modes produced by pairs of isotope
		impurities, 
	$|\Delta M/M| = 0.15$, $a = 10 a_{0}$ (solid lines) and $a = 4 a_{0}$ (dashed line). At $a = 4a_{0}$ the antisymmetric
	mode is delocalized and not shown here.} 
\label{fig:2_d_5}
\end{figure}

Equation (\ref{eq:twoimp2}) implies that at a sufficiently small distance between the isotopes $(a<L_{\ast})$ only one (even) 
mode $u^{[{\rm g}]}(-x)=u^{[{\rm g}]}(x)$ remains localized with $\omega>\omega_{\rm B}$ while the other, 
odd one $u^{[{\rm u}]}(-x)=-u^{[{\rm u}]}(x)$,  
is pushed into the phonon continuum and becomes delocalized. 
These features are shown in Fig. \ref{fig:IPR-2}, 
where we present the dependence of $\omega_{\rm loc}^{[{\rm g,u}]}$ and its $\mathcal{I}^{[{\rm g,u}]}$ on the distance 
$a$ between the isotopes. 
The corresponding Fourier components given by Eq. \eqref{eq:four} are presented in Fig. \ref{fig:2_d_5}. 

\subsection{Electron-phonon interaction} 

We present the electron Hamiltonian in the tight binding form
\begin{equation}
	H = - \frac{1}{2} \sum_{n}t_{n, n+1} \left(c_{n}^\dagger c_{n+1} + c_{n+1}^\dagger c_{n}\right)
\end{equation}
with the bond-dependent hopping (see Fig. \ref{chain})
$	t_{n, n+1} = t + \gamma^{\prime} (u_{n+1} - u_{n})$, with
$c_{n}, c_{n}^\dagger$ being the electron annihilation and creation operators (irrelevant spin degrees of freedom are 
not included). In the electron momentum basis 
the phonon-independent term $c_{k} = N^{-1/2} \sum_{n}e^{-i k a_{n}} c_{n}$ determines the one-dimensional electron dispersion
$\varepsilon_{k} = - t \cos{ka_{0}} $. 
The non-interacting part of the Hamiltonian is given by
\begin{equation}
	H_{0} = \sum_{k} \varepsilon_{k} c_{k}^{\dagger} c_{k} + \sum_{\alpha} \hbar \omega_{\alpha} b_\alpha^\dagger b_\alpha. 
\end{equation}
The electron-phonon interaction part can be written as
\begin{equation}
	H_{\rm e-p} = - \frac{ i \gamma^{\prime}}{\sqrt{N}} \sum_{k^{\prime},k} f_{k^{\prime},k} c_{k^{\prime}}^{\dagger} c_{k} \hat{u}_{q}, 
\end{equation}
where $\hat{u}_{q}$ is given by Eq.~(\ref{eq:four-uq}) 
with $q=k^{\prime}-k,$ and the form factor $f_{k^{\prime},k} = \sin{k^{\prime}a_{0}}-\sin{ka_{0}}.$ 
It is also convenient to rewrite the interaction potential in the form
that directly exploits $\phi_{q}^{\alpha}$ 
\begin{equation}\label{eq:Heph}
	H_{\rm e-p} = - i \gamma \sum_{k^{\prime},k\alpha} f_{k^{\prime},k}  
	c_{k^{\prime}}^{\dagger} c_{k} \left( b_{\alpha}^{\dagger} + b_{\alpha} \right) \sqrt{\frac{\omega_{\rm B}}{\omega_{\alpha}}} \phi_{q}^{\alpha},
\end{equation}
the electron-phonon interaction constant $\gamma = C(\hbar\omega_{\rm B}/\rho s^{2} L)^{1/2}/2\sqrt{2}$, 
where $C = \gamma^{\prime}a_{0}$ is  the deformation potential of the chain, 
$\rho=M/a_{0}$ is the mass density, and $L=Na_{0}$ is the chain length. 

\begin{figure}[h]
	\includegraphics[width=0.45\textwidth]{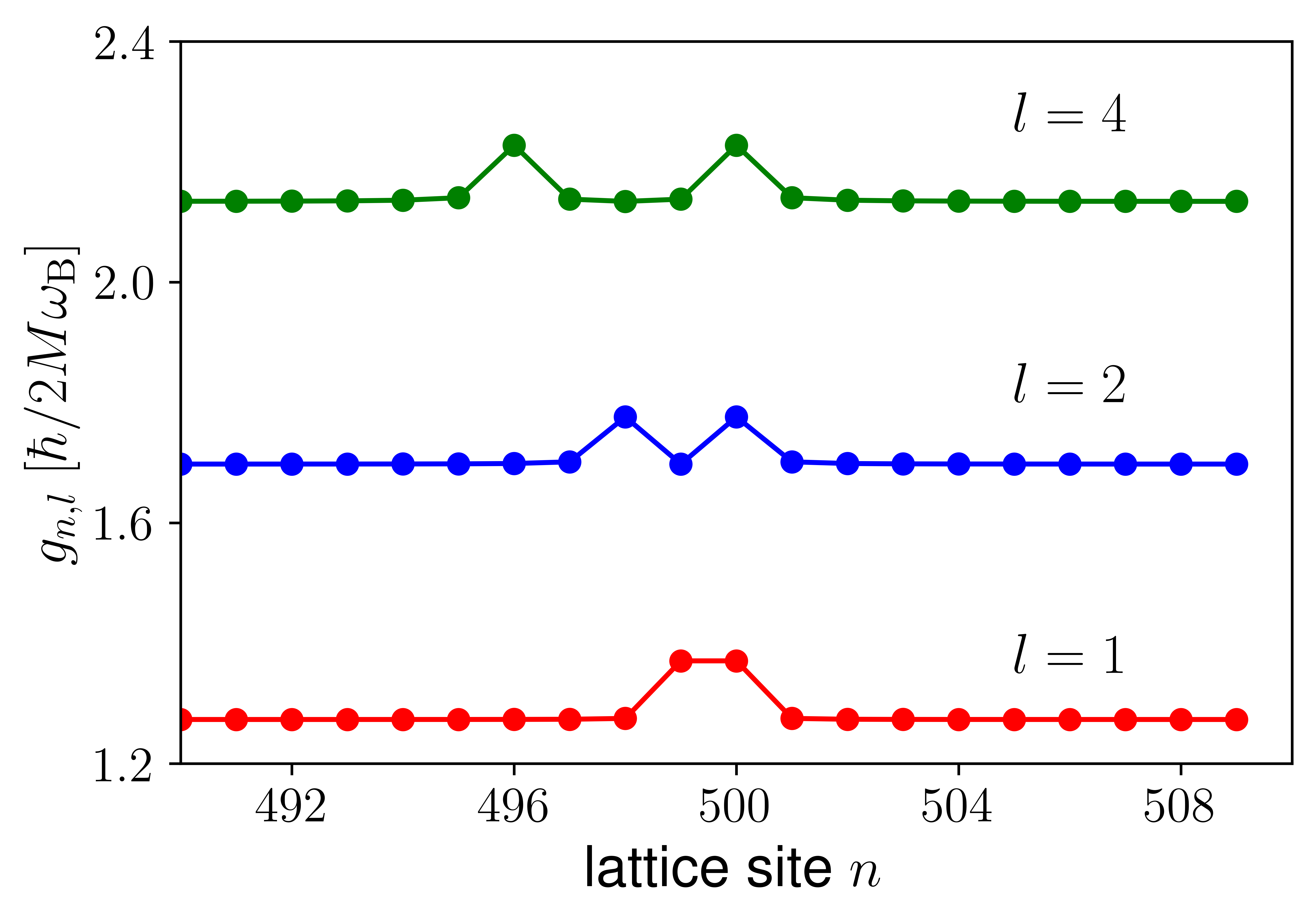}
	\caption{Correlation function $g_{n,l}$ for {different $l$ as marked near
	the plots.} The isotopic defect is located at $n=500$ and $|\Delta\, M|/M=0.15.$}
\label{correlator}	
\end{figure}

To explicitly confirm the effect of isotopic impurities on the quantum fluctuations of the 
intersite distances and corresponding electron hopping, we introduce the correlation 
function (cf. Eq. \eqref{unphin}):
\begin{equation}
	g_{n,l} = \langle (u_{n+l} - u_{n})^2 \rangle = 
 \sum_{\alpha} 
\frac{\hbar}{2 M \omega_{\alpha}}
\left( \phi_{n+l}^{\alpha} -\phi_{n}^{\alpha} \right)^2.	
\end{equation}
Figure \ref{correlator} demonstrates the dependence of $g_{n,l}$ (in the units of  $\hbar/2\omega_{\rm B}M$) 
on the lattice site $n$ in the vicinity of the isotope 
for different $l$. Though the localized phonon mode has a spatial size $L_\ast$, this scale is not reflected 
in the behavior of $g_{n,l}$. The latter is mainly determined by low-frequency 
phonons with small corrections appearing due to the mass defect $\Delta M$ at the isotope site. 
The increase in $g_{n,l}$ in the vicinity of the isotopic defect shows that the hopping integral fluctuates 
correspondingly and, thus can influence the electron motion. {However, as we will see below, the analysis 
of electron scattering cannot be reduced to using the only correlation function $g_{n,l}.$}

\begin{figure}[h]
	\includegraphics[width=0.45\textwidth]{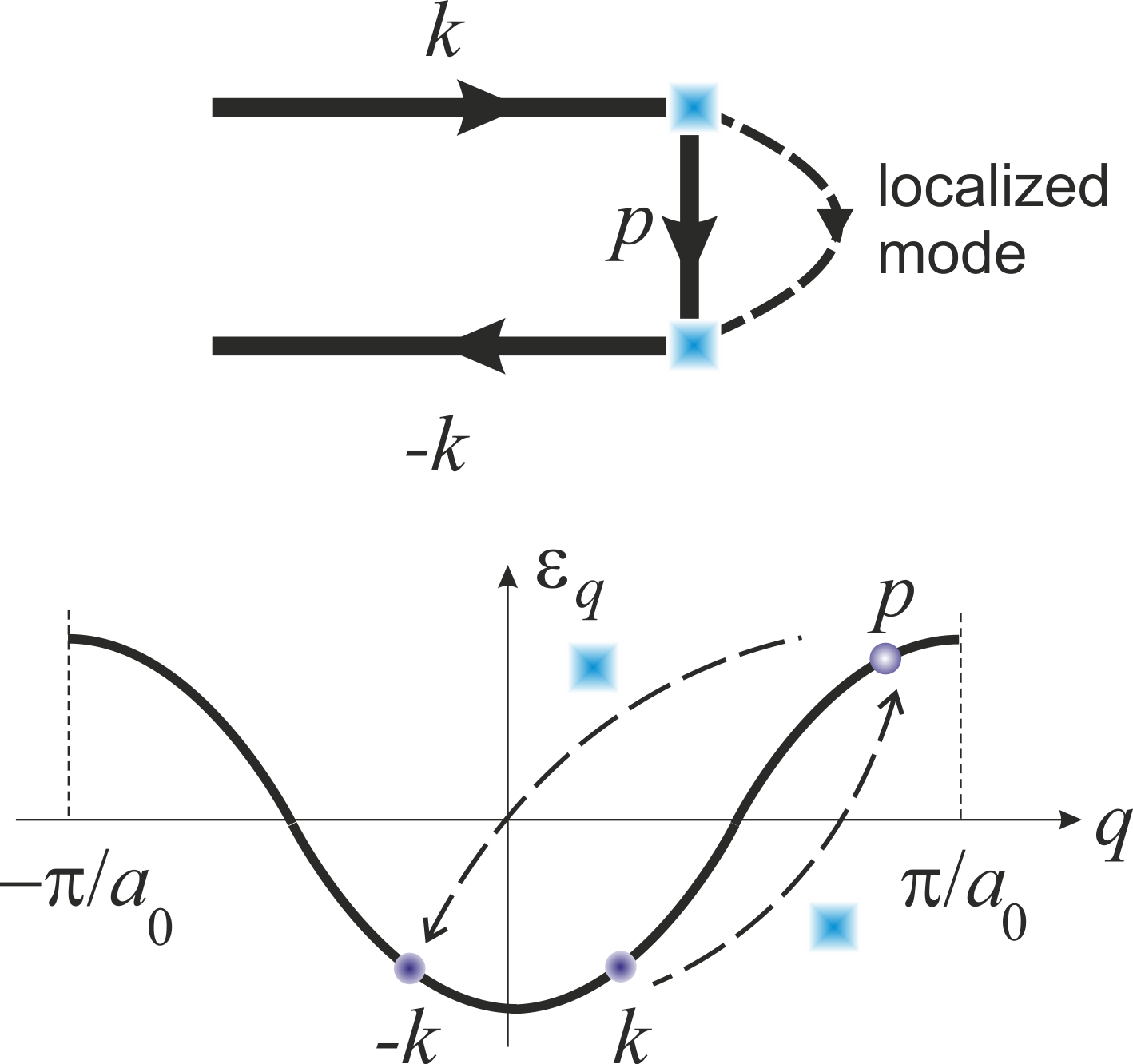}
	\caption{A backscattering process. Filled squares correspond to matrix elements of electron-phonon 
	coupling. Since the localized phonon mode has a spread in the momentum space, its 
	virtual emission and absorption causes elastic electron backscattering with momentum change $-2k.$ Notice that the main 
	contribution to the real scattering process is due to the virtual transition to the Brillouin zone boundary with 
	$|p|\approx\pi/a_{0}.$}
\label{backscattering}
\end{figure}

\section{Scattering and localization}

\subsection{Elastic scattering by localized modes} 

Aiming at analysis of localization we focus on the elastic electron scattering involving virtual phonon processes with 
the corresponding Feynman graph in Fig.~\ref{backscattering}.
This process appears in the second-order with respect to $\gamma$ and is due to the finite spread of phonon momentum in presence of isotopes. 
The effective potential for an electron interacting with quantum vibrations 
in second-order can be obtained following Refs. \cite{Luttinger,Kittel}  and is given by 
\begin{equation}
	\hat{V}
	= \frac{1}{2} \left[ H_{\rm e-p}, \hat{S} \right], 
\end{equation}
where $[,]$ stands for the commutator and operator 
$\hat{S}$ is determined from $[\hat{S}, H_{0} ] = H_{\rm e-p}.$ 
By taking the matrix elements of this commutator with respect to the eigenstates $(l,l^{\prime})$ of $H_{0}$ 
(such as $|k 1_{\alpha}\rangle \equiv c_{k}^{\dagger}b_{\alpha}^{\dagger}|0\rangle,$ where $|0\rangle$ is the vacuum
state), we obtain
\begin{equation}
	S_{l,l^{\prime}} = \frac{H_{\rm e-p}^{l,l^{\prime}}}{E_{l^{\prime}} - E_{l}}.
\end{equation}
To analyze the elastic scattering we first consider the matrix elements of $\hat{V}$ between 
$|k0\rangle$ and $|k^{\prime} 0\rangle$ phononless states. 
The explicit expression for this matrix element, $V_{k^{\prime},k}$, is
\begin{eqnarray}
	\label{eq:VKKK}
	&&	V_{k^{\prime},k} = \frac{1}{2} \sum_{p,\alpha}
	\langle k^{\prime}0|  H_{\rm e-p}  | p 1_{\alpha} \rangle
	\langle p 1_\alpha | H_{\rm e-p}  | k0 \rangle \times \notag\\
	&&	\left(
	\frac{1}{\varepsilon_{k} - \varepsilon_{p} - \hbar\omega_{\alpha}} + \frac{1}{\varepsilon_{k^{\prime}} - \varepsilon_{p} - \hbar\omega_{\alpha}}
	\right).
\end{eqnarray}
At small electron momenta the electron velocity $v_{e} = (d\varepsilon_{k}/dk)/\hbar= k t a_{0}^{2}/\hbar$ 
and effective mass $m$ with $1/m=(d^{2}\varepsilon_{k}/dk^{2})/\hbar^{2}=ta_{0}^2/\hbar^{2}.$ 
{To avoid decoherence processes related to the emission of real phonons and suppressing the electron localization,  
we note that the no-emission condition $v_{e}<s$ restricts momentum 
to $k<k_{\rm em},$ where $k_{\rm em}=\hbar\omega_{\rm B}/2ta_{0}.$}
Notice that the condition $t\left(k_{\rm em}a_{0}\right)^{2}\sim \hbar^{2}\omega_{\rm B}^{2}/t\ll\hbar\omega_{\rm B}$
prohibits the emission of real localized modes also. 
{Using $H_{\rm e-p}$ from Eq.~(\ref{eq:Heph}) 
we keep in the sum of Eq.~ (\ref{eq:VKKK}) only the modes corresponding to the localized states of Eq.~ (\ref{eq:eqloc}) 
with frequency $\omega_{\alpha} = \omega_{\rm loc}.$ 
For a low concentration of single isotopes with a small mass defect, the Fourier components of other, delocalized, phonon modes
with the frequencies lying within the phonon band spectrum, do not acquire a width sufficient to contribute to the sum in Eq.~(\ref{eq:VKKK}).}
Thus, we arrive at the matrix element of the electron scattering by a single isotope impurity: 
\begin{eqnarray}\label{eq:Gkk_analit}
	&& V_{k^{\prime},k}^{[1]} = - \frac{\gamma^2}{t} \Gamma_{k^{\prime},k}^{[1]}, \\	
	&&\Gamma_{k^{\prime},k}^{[1]}=\frac{1}{2}\sum_{p}f_{k^{\prime},p}f_{p,k} 
	\phi_{k^{\prime}-p}^{\alpha}\phi_{p-k}^{\alpha}
	\left(\frac{1}{\Delta(p,k)} + \frac{1}{\Delta(p,k^{\prime})}\right), \notag \\
&&\Delta(p,q) = {2 k_{\rm em} a_{0} - \cos{pa_{0}} + \cos{qa_{0}}}. \notag	
\end{eqnarray}

\begin{figure}[h]
	\includegraphics[width=0.5\textwidth]{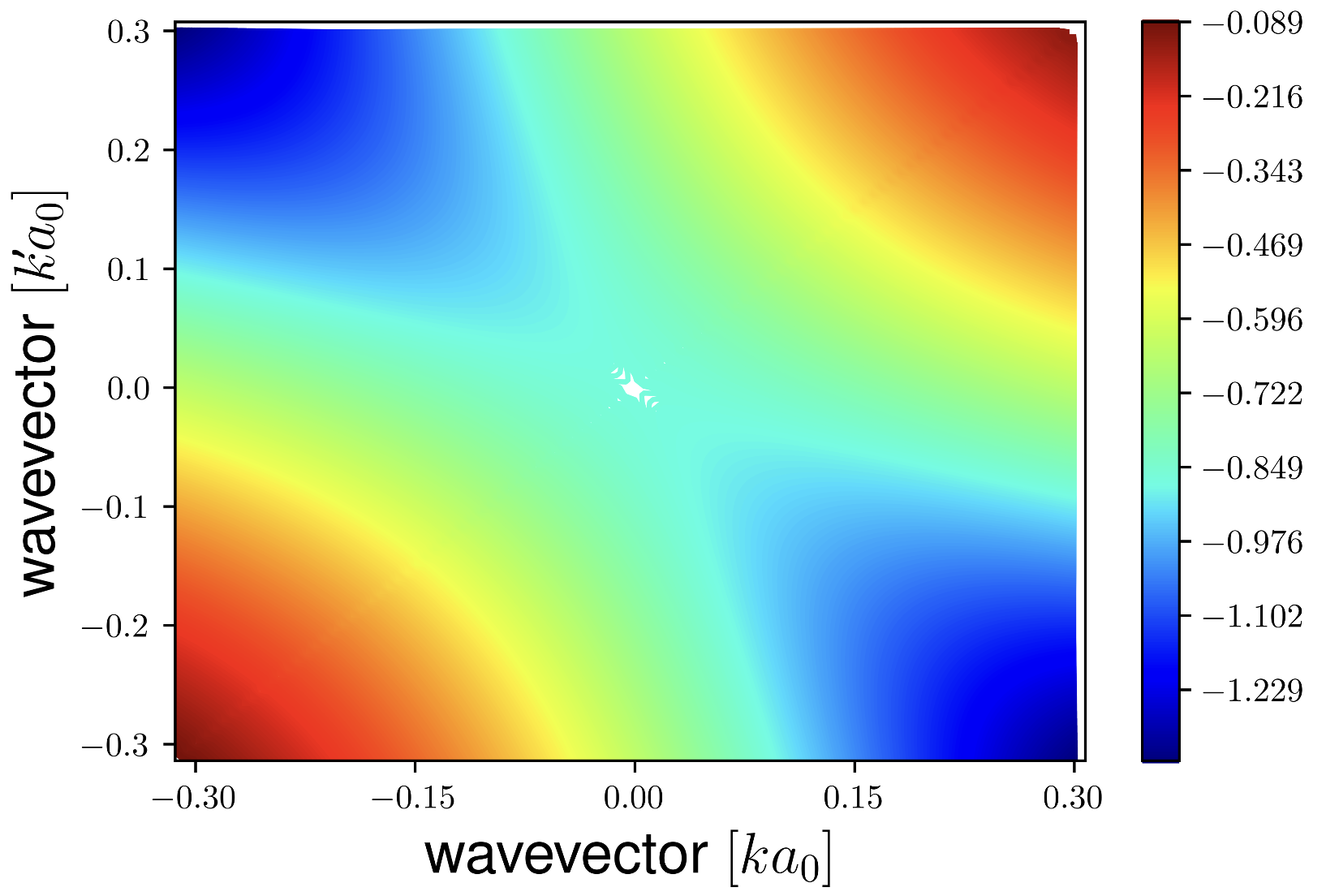}
	\caption{Logarithm of matrix element $\log_{10}\Gamma_{k^{\prime},k}$ for $\Delta M/M =-0.15.$} 
\label{fig:Vef}
\end{figure}

\begin{figure}[h]
	\centering
	\includegraphics[width=0.5\textwidth]{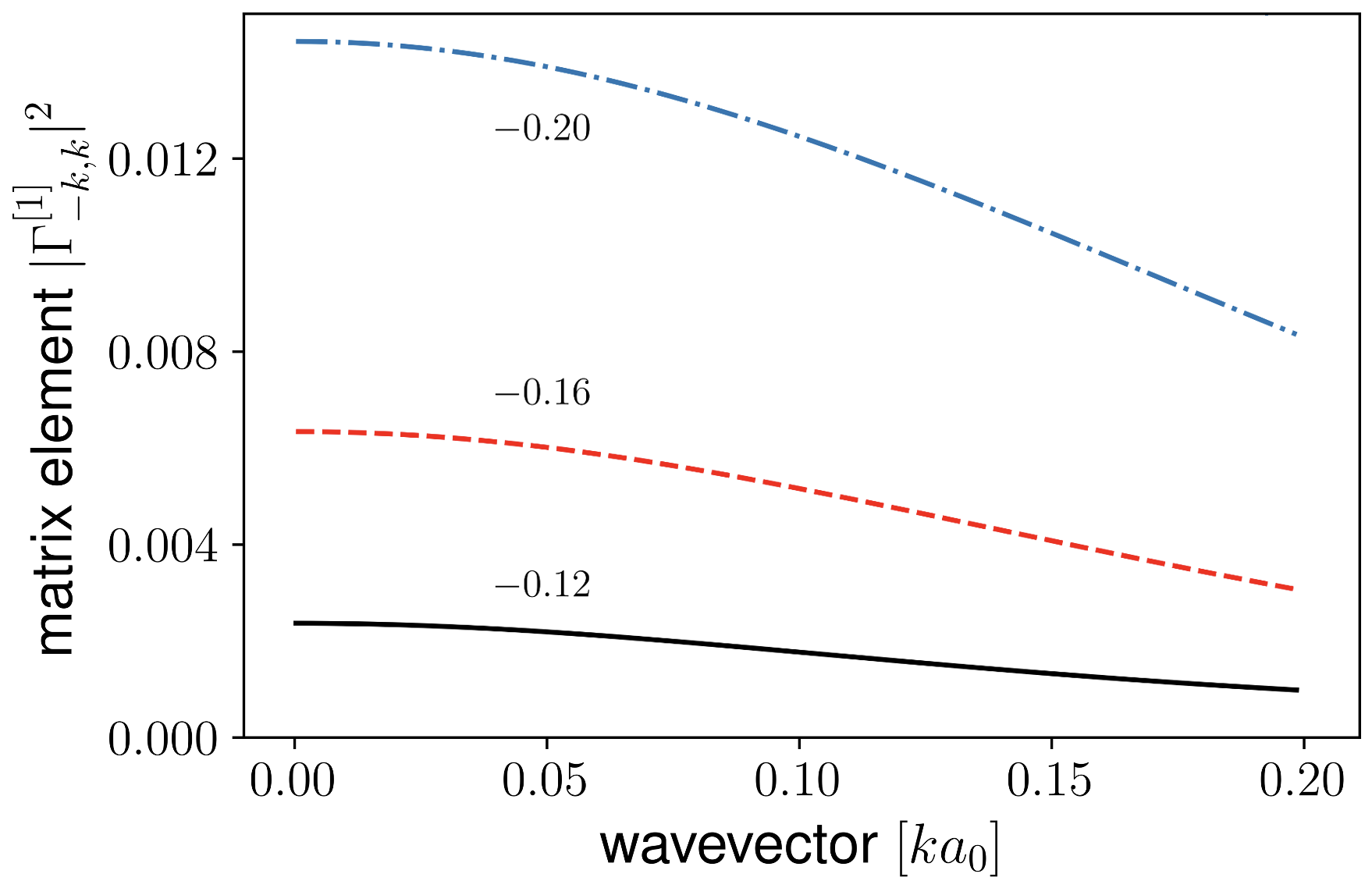}\\
	\caption{$|\Gamma_{-k,k}^{[1]}|^{2}$ for $\Delta M/M$ shown. 		 
	Fit according to Eq. \eqref{eq:Vkk_analit} is not shown 
	since it almost coincides with the numerical curves. The parameter $k_{\rm em}a_{0} = 0.2$.}
\label{fig:Mpp}
\end{figure}

The plot of $\log_{10}\Gamma_{k^{\prime},k}$ corresponding to $V_{k^{\prime},k}^{[1]}$ 
presented in Fig. \ref{fig:Vef} demonstrates 
{that the scattering matrix element is relatively large only in the $ q\lesssim\,L_{\ast}^{-1}$  range.  
The resulted calculated weakness} of backscattering is due to two factors: a small allowed momentum transfer with  
$q = 2k \lesssim\,L_{\ast}^{-1}$, arising from the $\phi_{p}^{\rm loc} \phi_{p+q}^{\rm loc}$, see Eq.(\ref{eq:eqloc}), 
and small $f_{k,p}$ at small $k$
and $p\approx q_{\rm B}$. By using the condition  
 $k_{\rm em}a_{0}\ll\,1$ and expression for $\phi_{q}^{\rm loc}$ in Eq. \eqref{eq:eqloc}, we obtain:
\begin{equation} \label{eq:Vkk_analit}
|V_{-k,k}^{[1]}|^{2} =
	\left(\frac{\gamma^2 a_{0}^{2}}{2t L_{\ast}^{4}} \right)^{2}
	\frac{1}{\left(k^{2} + L_{\ast}^{-2} \right)^{2}}. 
\end{equation}
In the $kL_{\ast} \ll 1$ limit  $|V_{-k,k}^{[1]}|^{2}$ behaves as $(\Delta M/M)^{4}$ and decreases as $(\Delta M/M)^{8}/k^{4}$ 
at $k \gg 1/L_{\ast}.$ {Equations (\ref{eq:Gkk_analit}) and (\ref{eq:Vkk_analit}) present the key result of this paper: 
Quantum isotopic impurities characterized by localized finite-frequency phonon modes can produce elastic electron backscattering.
The corresponding numerically calculated $|\Gamma_{-k,k}^{[1]}|^{2}$ is presented in Fig. \ref{fig:Mpp}.}

\begin{figure}[h]
	\includegraphics[width=.5\textwidth]{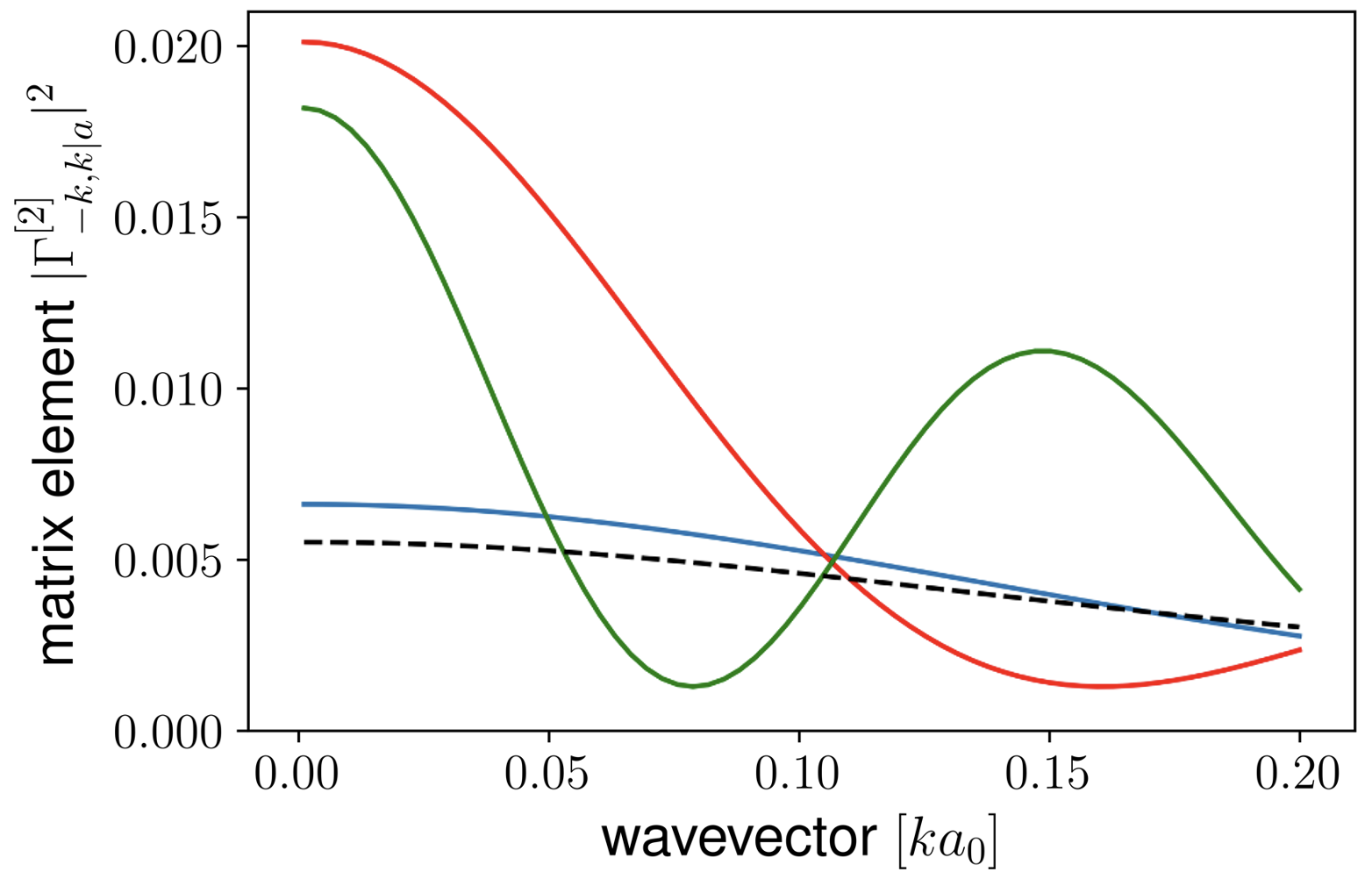}		
	\caption{Backscattering probability by two-isotope phonon states. Here $\Delta M/M = -0.15$, $k_{\rm em}a_{0} = 0.2$. 
		The distance between isotopes: $a = 2 a_{0}$ (blue),  $a = 10 a_{0}$ (red), $a = 20a_{0}$ (green). 
	The dashed line corresponds to a single isotope.} 
\label{fig:2_dist}
\end{figure}

{Having demonstrated possible elastic scattering by a single isotopic impurity, we can study how this general 
approach can be applied for two closely related impurities, where interference of scatterings by 
odd and even vibrational modes plays a qualitative role, and  see the characteristic features of this 
scattering.} The matrix element of backscattering 
by two-isotope states, corresponding to Figs. \ref{fig:IPR-2} and \ref{fig:2_d_5},
has the form:
\begin{eqnarray}
&&V_{-k,k}^{[2]} = - \frac{\gamma^{2}}{t} \Gamma^{[2]}_{-k,k|a}, \\
&& \Gamma^{[2]}_{-k,k|a} = \sum_{p,[{\rm s}]}
\frac{1}{\Delta(p,k)}
f_{-k,p} f_{p, k} 
\phi_{-k - p}^{[{\rm s}]}  \phi_{p- k}^{[{\rm s}]},
\notag
\end{eqnarray}
and shows interference of the scattering processes. 
The corresponding probabilities are presented in Fig. \ref{fig:2_dist}. 
It is worth mentioning an increase in the scattering probability when two 
isotopes are located at $a \lesssim 4 L_{\ast}$, indicating a stronger scattering. 

\subsection{Low density of impurities: possible single-impurity localization and free path} 

As we demonstrated in the previous subsection, at small isotopic defects concentrations, 
a disordered chain can be presented as a 
randomly spaced ensemble of elastic scatterers, corresponding to scattering by a single and pairs of isotopic impurities. 
{We assume a sufficiently small concentration of 
isotope impurities $\nu_{\rm is}\ll\,|\Delta M|/Ma_{0}$ such 
that they behave as independent single scatterers located at uncorrelated points $x_{i}.$
The resulting small concentration of isotope pairs of the order of $\nu_{\rm is}^{2}L_{\ast}\ll\,\nu_{\rm is}$ 
does not modify the localization-related results.}

We begin with the quantum mechanical analysis of the electron free path in this system,  
where the finite free path indicates localization behavior and assuming that the main 
effect is due to single scatterers. 
In this analysis we use the one-dimensional free space electron Green's function in the form:
\begin{equation} \label{eq:GF}
G_{E}^{(\pm)}(x,x^{\prime})=\pm\frac{im}{k\hbar^{2}}\exp\left(\pm ik \left|x-x^{\prime}\right|\right),
\end{equation}
where $E=\hbar^{2}k^{2}/2m.$ Using the Green function from Eq.~(\ref{eq:GF}) 
we obtain in the first Born approximation for scattering by a single isotope impurity located at a point $x_{i}$:
\begin{eqnarray}\label{Psi_full}
&&\Psi = e^{ik(x-x_{i})} + r(k) e^{-ik(x-x_{i})}, \\
&& r(k) = -\frac{i m}{\hbar^{2} k} \widetilde{V}^{[1]}_{-k,k}, \nonumber
\end{eqnarray}
where $x<x_{i},$ $|x-x_{i}|\gg\,L_{\ast},$ $\widetilde{V}^{[1]}_{-k,k} = L V^{[1]}_{-k,k}$ 
and one expects $|r(k)|\ll\,1$ as the approximation validity condition.  
The sign of $\widetilde{V}^{[1]}_{-k,k}$ is negative corresponding to an attractive potential of the width $L_{\ast},$
depth of the order of $t\times(\hbar\omega_{\rm B}/M s^{2})\times(a_{0}/L_{\ast})^{3},$ 
and the binding energy $\epsilon_{\rm loc}$
of the order of $t\times(\hbar\omega_{\rm B}/M s^{2})^{2}\times(a_{0}/L_{\ast})^{4}.$ 
{Here we used relations $Ms^{2}=\kappa a_{0}^{2}$
to scale the energy in dimensionless units and $t\sim\,C=\gamma^{\prime}a_{0}$ by 
assuming that the intersite hopping integral is sufficiently modified 
at the modulation in the interatomic distance of the order of  $a_{0}$ (cf. Eq. (\ref{eq:Heph})).}
 
In the positive energy domain, the mean free path due to back scattering by isotopes 
can be estimated as {$\ell(k) = \nu_{\rm is}^{-1}|r(k)|^{-2}\gg\,\nu_{\rm is}^{-1},$ 
corresponding to {the Anderson} localization 
at the spatial scale of the order of $\ell(k).$} 
Indeed $|r(k)|$ is very small almost in the entire interval of $k<k_{\rm em}$ at the energies 
$\varepsilon_{k}+t>10^{-2}(\Gamma^{[1]}_{-k,k})^{2}\left(\hbar^{2}/Ma_{0}^{2}\right)\times\left(t/Ms^{2}\right).$
At smaller energies, $|r(k)|$ approaches 1 and more detailed scattering approaches \cite{Halperin,Azbel} have to be applied.
To characterize the scattering, one can introduce dimensionless parameters 
$k_{\rm em}L_{\ast}\sim \hbar\omega_{\rm B}/t\times(M/|\Delta M|)$  and 
$\omega_{\rm B}\tau\sim\ M/|\Delta M|,$ where, {due to the condition $v_{e}<s,$ the parameter $\tau = L_{\ast}/s$ represents the 
upper limit of the scattering time.} Although $k_{\rm em}L_{\ast}$ can be either smaller or greater than 1, the product $\omega_{\rm B}\tau$ is always
much larger than 1. This implies that scattering by isotopic impurities is an adiabatic process with low probability due
to averaging of the intersite hopping during the scattering leading as a result to the adiabatic limit for the binding 
energy $\epsilon_{\rm loc}\ll\hbar\omega_{\rm B}.$

\subsection{Electron localization} 

Here we concentrate on electron localization by using the result of previous subsections
on mapping of isotopic impurities in chains on 
weak scatterers of electrons. As it is was shown in Ref. \cite{Berry}, 
one-dimensional arrays of random scatterers lead to the localization of light waves with the same  
approach valid for the localization of electron waves.

\begin{figure}[h]
	\includegraphics*[width=0.45\textwidth]{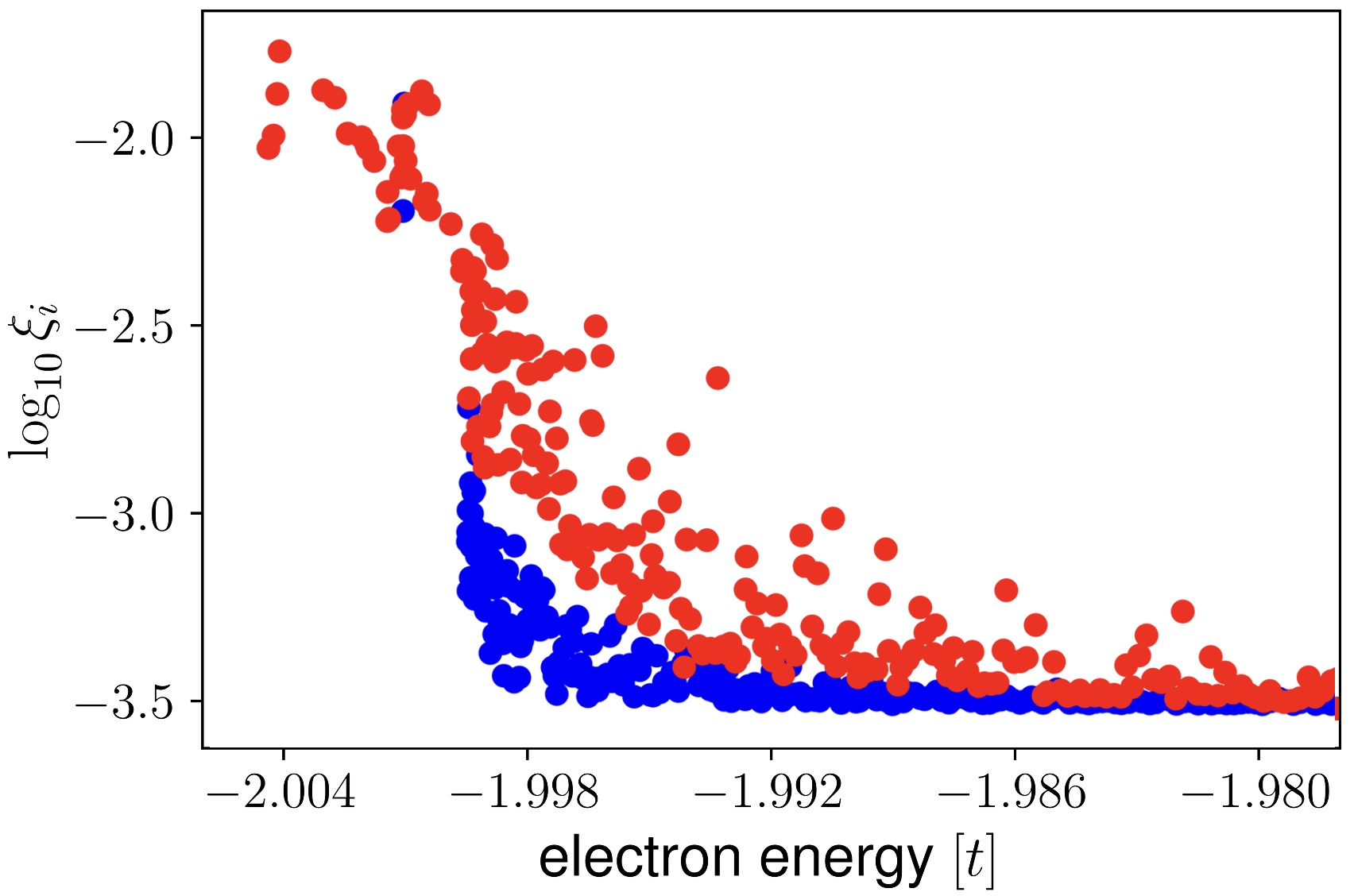}\\
	\caption{{The ${\rm log}_{10}\xi_{i}$ of inverse participation ratio for two low-density distribution of isotopic impurities.
	The concentration of impurities $\nu_{\rm is}=0.002$ (blue circles) and $\nu_{\rm is}=0.009$ (red circles). 
	The impurity width $L_{\ast}=5 a_{0}$ corresponding to $\Delta M/M = -0.1$ and the effective potential 
	amplitude is $\tilde{V}=-0.01.$ The hopping integral $t=1.$ }}
\label{fig:logipr}
\end{figure}

Several approaches including the Lyapunov exponents analysis (e.g., \cite{Gurevich2011}) 
and random matrix theory (e.g., \cite{Beenakker1997}) can be used for studies of wave localization in  random potentials.  
We will use a direct numerical calculation proving  electron localization by producing  a random potential corresponding to 
isotopic scatterers and calculating electron eigenstates in this potential. To study the localization 
we consider a normalized electron wavefunction $\psi_{i}(n),$ on a $N\gg\,1-$sites one-dimensional lattice, 
where $n=1,\ldots,N$ enumerates the lattice sites and $i$ enumerates the electron states with the energies $\varepsilon_{i}.$ 
The probability density distribution for the state $i$ is characterized by the corresponding inverse participation ratio:
\begin{equation}\label{eq:iprelectron}
\xi_{i}=\sum_{n}|\psi_{i}(n)|^{4}.
\end{equation}

\begin{figure}[h]
	\includegraphics*[width=0.45\textwidth]{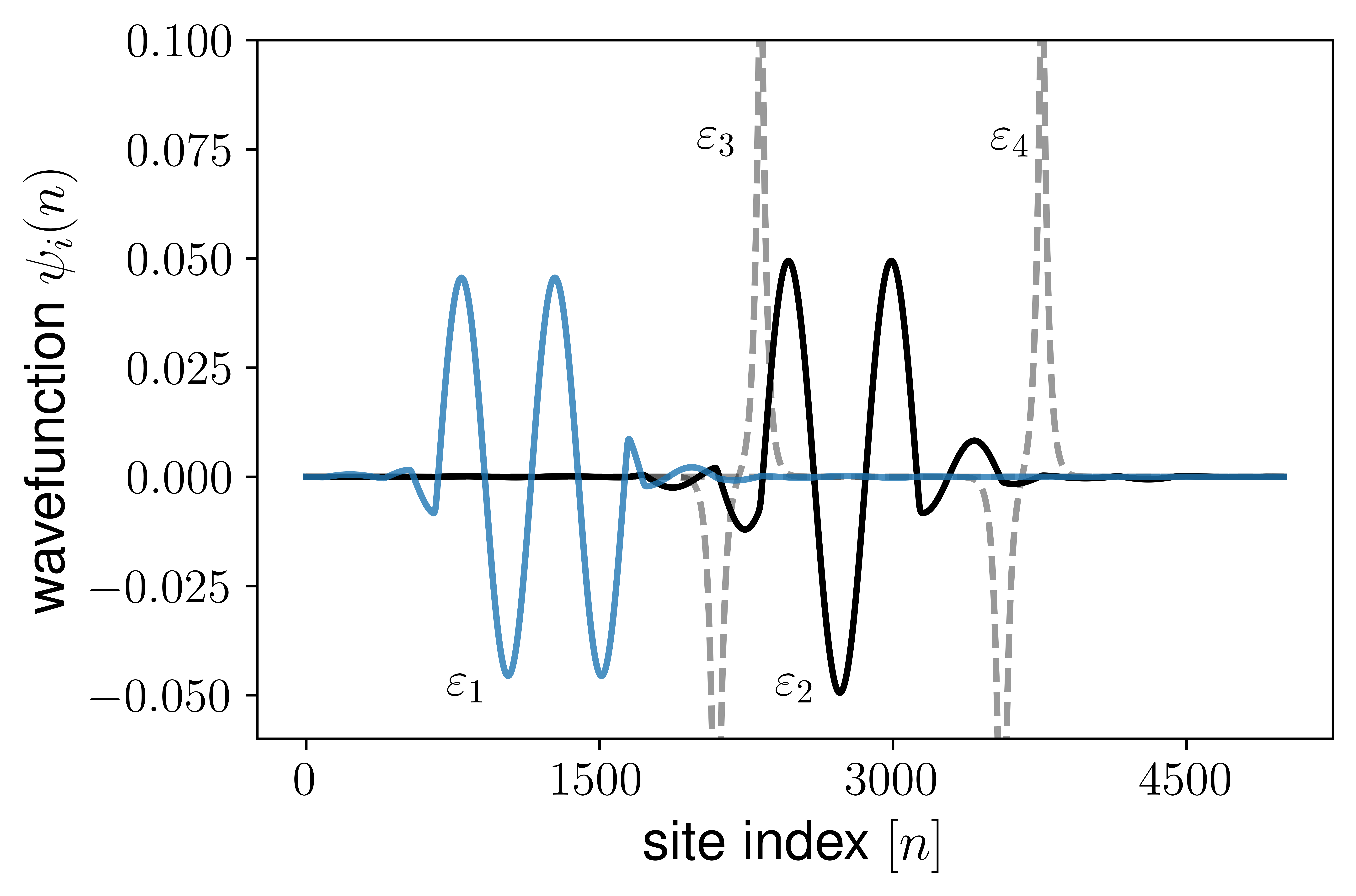}\\
	\caption{{Electron wavefunctions in different states with energies presented near the plots. Dashed lines: states  in the impurity band
	with $\varepsilon_{4}+2t = -5\times 10^{-6}t$ and $\varepsilon_{3}+2t = -1.4\times 10^{-3}t.$  
	Solid lines:  states in the conduction band described by the Anderson localization with $\varepsilon_{2}+2t = 2.4\times 10^{-4}t$ and $\varepsilon_{3}+2t = 3.4\times 10^{-4}t.$ .
	}} 
\label{fig:locprob}
\end{figure}

{The results for ${\rm log}_{10}\xi_{i}$ presented in Fig. \ref{fig:logipr} demonstrate the formation of two groups of localized states. The first
group corresponds to the impurity band formed below the bottom of the conduction band with $\varepsilon_{i}<-2t$ \cite{Lifshitz_book}. 
The second group corresponds to the Anderson localization of the states inside the conduction band with $\varepsilon_{i}>-2t.$ 
We notice that, as expected, the states become delocalized at the energies corresponding to $k>1/L_{\ast},$ 
where the backscattering probability is strongly suppressed, according to 
{Eq. (\ref{Psi_full}) and Eq. (\ref{eq:Vkk_analit}), where $V_{-k,k}^{[1]}$ 
behaves as ${1}/\left(k^{2} + L_{\ast}^{-2}\right),$ rapidly decreasing with the 
electron energy.} Further, in Fig. \ref{fig:locprob} we present several typical wavefunctions $\psi_{i}(n)$ 
demonstrating that the {Anderson} localization occurs on a spatial scale much 
larger than the distance between the isotopic impurities.} 

Finally, we emphasize that our approach is qualitatively 
different from that proposed in Refs. \cite{Berezin1982,Berezin1984a,Berezin1984b,Berezin1987}.
{The analysis in these papers is based on the fact that the electron self energy  due 
to electron-phonon coupling, producing bandgap renormalization, is proportional to the 
dependent on the atomic masses mean inverse phonon frequency \cite{Kittel}. Thus, clusters of isotopic impurities 
with sufficiently large size and concentration, renormalizing locally the electron bandgap, 
can produce effective ``potentials'' able to localize sufficiently heavy carriers. Our approach 
is unrelated to the bandgap renormalization and electron localization occurs do to elastic 
scattering by randomly distributed localized phonon modes.}

\section{Conclusions and outlook} 

We studied the localization of electrons due to a weak {isotopic disorder producing  
randomly spatially distributed localized phonon modes} in one-dimensional 
systems described by a tight-binding model. Elastic electron scattering 
{involving nonlocal processes of} virtual 
emission and absorption of these modes is presented in the form of scattering by 
{a weak} potential with a strongly energy-dependent reflection probability. 
This randomness, related to the virtual phonon emission and absorption processes 
leads to the {Anderson} localization 
with the localization length dependent on the electron energy. Investigations of more complex realizations including the formation 
of random weak coupling polarons due to a strong isotopic disorder and possible effects of lattice nonlinearity 
(see, e.g. Ref. \cite{Savin}) are of interest and will be a topic of future research. 

To conclude the discussion of the effect of isotopic mass on the {electron position in systems with different isotopes that
emphasizes its general character,} we mention that this effect 
is related to the puzzling dipole moment of the HD molecule \cite{Trefler}, where different amplitudes of 
zero-point vibrations of hydrogen and deuterium ions lead to a shift of the electron probability density between these atoms 
resulting in the formation of a small dipole moment \cite{Trefler,Blinder}.

\section{Acknowledgements} 

The work of E.S. is supported through Grants
No. PGC2018-101355-B-I00 and No. PID2021-126273NB-I00 funded by MCIN/AEI/10.13039/501100011033 and by
the ERDF “A way of making Europe”, and by the Basque Government through Grant No. IT1470-22. E.S. is grateful to
C. Draxl, N. Protik, J.E. Sipe, and V.A. Stephanovich for valuable discussions.

\end{document}